\documentclass[aps, prb,a4paper,showpacs,reprint]{revtex4-1}
\pdfoutput=1
\usepackage{graphicx}
\usepackage{color}
\usepackage{comment}
\usepackage{array}
\usepackage{amsmath}
\usepackage{upgreek}
\usepackage{color,soul}

\definecolor{gray}{gray}{0.6}
\setulcolor{yellow}

\newcolumntype{C}[1]{>{\centering\let\newline\\\arraybackslash\hspace{0pt}}p{#1}}

\begin{document}

\title{Thermal transport in nanocrystalline graphene investigated by approach-to-equilibrium molecular dynamics simulations}
\author{Konstanze R. Hahn}
\email[e-mail: ]{konstanze.hahn@dsf.unica.it}
\author{Claudio Melis}
\author{Luciano Colombo}
\affiliation{Dipartimento di Fisica, Universit\`a di Cagliari\\ Cittadella Universitaria, I-09042 Monserrato (Ca), Italy}

\begin{abstract}
Approach-to-equilibrium molecular dynamics simulations have been used to study thermal transport in nanocrystalline graphene sheets. Nanostructured graphene has been created using an iterative process for grain growth from initial seeds with random crystallographic orientations. The resulting cells have been characterized by the grain size distribution based on the radius of gyration, by the number of atoms in each grain and by the number of atoms in the grain boundary. Introduction of nanograins with a radius of gyration of 1 nm has led to a significant reduction in the thermal conductivity to 3\% of the value in single crystalline graphene.
Analysis of the vibrational density of states has revealed a general reduction of the vibrational intensities and broadening of the peaks when nanograins are introduced which can be attributed to phonon scattering in the boundary layer.
The thermal conductivity has been evaluated as a function of the grain size with increasing size up to 14 nm and it has been shown to follow an inverse rational function.
The grain size dependent thermal conductivity could be approximated well by a function where transport is described by a connection in series of conducting elements and resistances (at boundaries).
\end{abstract}

\maketitle

\section{Introduction}
Graphene is a material of great interest from a technological point of view owing to its high carrier mobility and optical transparency. Furthermore, it provides excellent physical strength and is chemically inert towards many environmentally ubiquitous species.\cite{Geim2007,Lee2008,Nair2008,Mayorov2011}
Such properties paved the way for graphene's career in applications such as field effective transistors, Schottky junction diodes or as transparent and flexible displays in various electronic devices.\cite{Lin2010,Bae2010,Britnell2012,Liu2013,Kalita2013}

Miniaturization of electronic devices is an important topic in recent research. A major issue for such devices is the required heat dissipation which becomes more and more important with decreasing dimensions. Crystalline graphene is a promising material in this regard. It inheres both high electronic carrier mobility and excellent thermal conductivity which additionally provides the function as heat dissipator.

Fabrication of single-crystalline graphene sheets, however, is not trivial.
Common fabrication methods include epitaxial film growth and chemical vapor deposition (CVD), where the latter is promising in particular for large-scale production. In all methods, defects are introduced in the crystalline structure resulting from limitations of the kinetic properties in the growth process and defects of the substrates. These defects limit the size of single-crystalline domains rendering the graphene sheets rather polycrystalline. The introduction of grain boundaries and defects in such polycrystalline graphene can drastically change electronic and thermal transport properties of the system.
In fact, previous experimental measurements of the thermal conductivity in graphene resulted in values ranging from 600 to 10000 W/mK.\cite{Balandin2011} This wide range in the value of thermal conductivity demonstrates its high sensitivity to various process conditions, sample size, and structural features including defects and grain boundaries.

Characterization and control of the structural properties of polycrystalline graphene, such as the grain size, and their effect on the electrical and thermal transport properties are thus of major interest in research of graphene in electronic and thermoelectric devices.
Regarding the effect of environmental parameters, a significant difference in the thermal conductivity has been found for suspended ($\sim$1600 W/mK)\cite{Chen2011} and SiO$_2$ supported (600 W/mK)\cite{Seol2010} graphene.
Furthermore, thermal conductivity of CVD produced graphene has been shown to be reduced by ~30\% when wrinkles are introduced in the structure.\cite{Chen2012} A systematic investigation of the size effect of single-crystalline domains on the thermal conductivity of graphene based on experiments, however, is still matter of investigation.

Due to the complexity of generating trustworthy polycrystalline simulation cells, little research has been done so far on the theoretical investigation of the properties of polycrystalline graphene.
Applying atomistic simulations it has been shown recently that the mechanical strength of graphene is reduced by 50\% when grain boundaries in form of polycrystals are introduced.\cite{Kotakoski2012}
Furthermore, equilibrium molecular dynamics (EMD) simulations have been carried out where the thermal conductivity has been described as a function of the polycrystalline grain sizes.\cite{Mortazavi2014}

In this work, in addition to the grain size dependent thermal conductivity of nanocrystalline graphene, the morphology of crystalline domains has been analyzed in detail. Furthermore, the phonon spectrum has been extensively investigated based on the vibrational density of states (VDOS) and the accumulation of bulk thermal conductivity.
First, we developed a model to create nanocrystalline graphene sheets with different grain sizes and determined their particle size distributions. 
Thermal transport properties of these systems have then been studied by approach to equilibrium molecular dynamics (AEMD) calculations.\cite{Lampin2013,Melis2014a}
Two specific systems with a grain size of 1 and 2.5 nm, respectively, have been chosen for a detailed elaboration of the thermal transport including the extrapolation of the bulk thermal conductivity and the analysis of the phonon spectrum based on the accumulation function of the thermal conductivity.
Furthermore, the vibrational density of states (VDOS) in nanocrsytalline graphene has been evaluated and compared to the  VDOS of pristine graphene.
Finally, the thermal conductivity of systems with varying average grain size from 0.7 to 14 nm has been investigated showing increasing thermal conductivity with increasing grain size.
The thermal conductivity has been approximated as rational functions both of the grain size and the number (concentration) of atoms in the grain boundaries.

\section{Methods}

\subsection{Generation of nanocrystalline graphene}
In this study, thermal transport in extended (bulk) nanocrystalline graphene has been studied using atomistic simulations.
Computer generation of nanostructured materials, however, is not trivial, in particular, when periodic boundary conditions have to be met as it is the case in extended bulk systems. In analogy to experiments, production of such nanostructured systems can be realized by bottom-up or top-down methods. Top-down methods include the creation of a homogeneous amorphous or crystalline simulation cell which is used as a matrix material. Manual manipulation by scissoring and insertion of different seeds and structures is then used to create nanostructured materials. This method has been used, for example, for the creation of nanocrystalline SiGe alloys.\cite{Melis2014b} In contrast, in bottom-up methods nanostructures are constructed from single atoms or small building blocks (for example single-crystalline domains). Two-dimensional (2D) graphene sheets simulated here have been simulated using the bottom-up approach.
\begin{figure}[tb]
\centering
   \includegraphics[width=.4\textwidth]{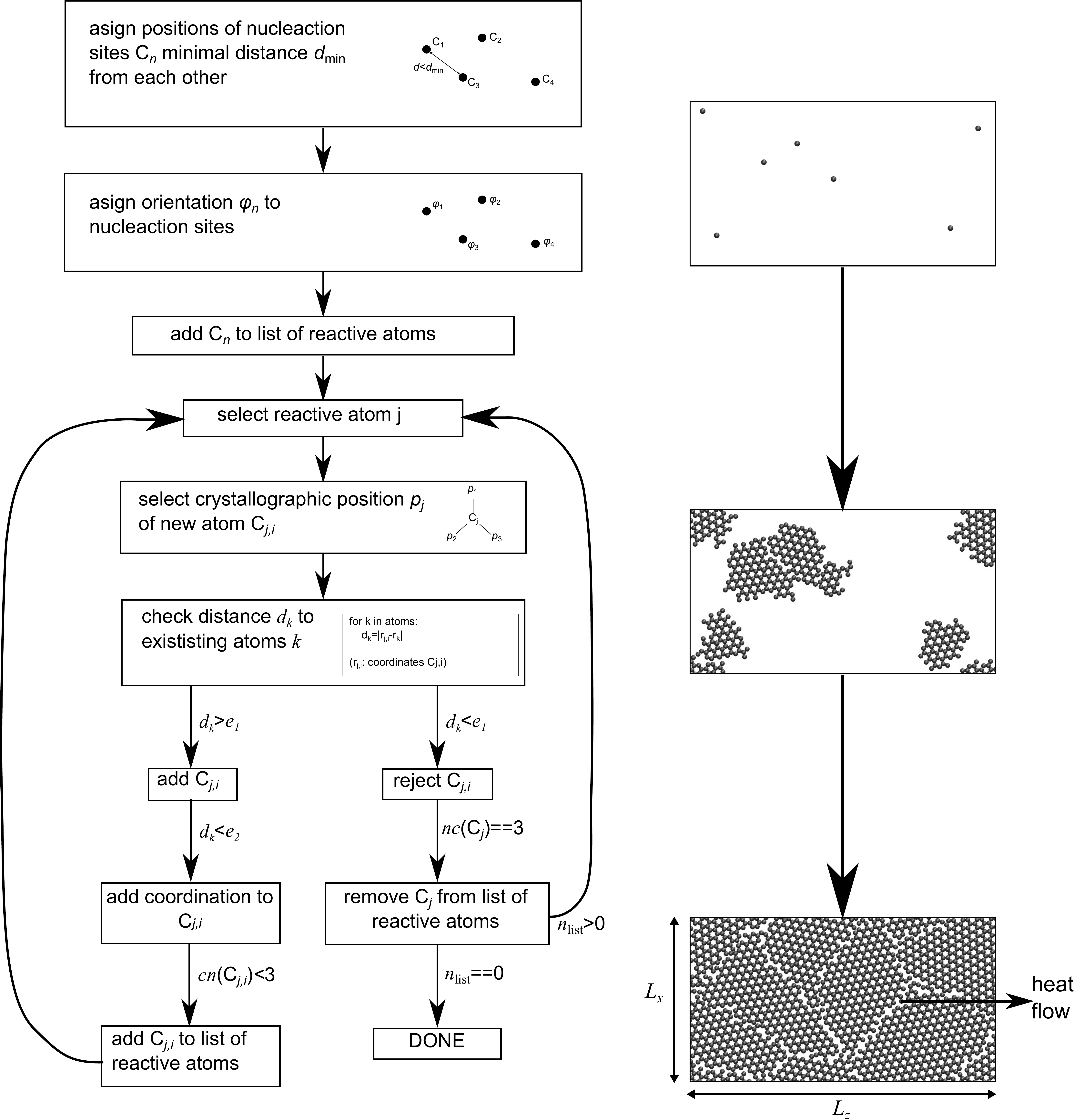}
       \caption{
       Schematic of the simulation cell and the algorithm developed for creation of nanocrystalline graphene. In the above scheme, $cn$($C_{j,i}$) indicates the coordination number of the respective atom $i$.
       }
       \label{fig:algo}
\end{figure}

An algorithm has been developed where, for a given rectangular simulation cell with width $L_x$ and length $L_z$ (orthogonal to and in direction of the thermal transport, respectively), a specified number of differently oriented crystalline grains has been placed inside the simulation box (Figure~\ref{fig:algo}).
Initial nucleation sites have been placed randomly in the simulation cell with the constraint that neighboring seed atoms have a distance $d_\mathrm{min}$ between each other of at least one fifth of the ideal average grain size ($d_\mathrm{min}=\frac{d_\mathrm{ave}}{5}=\frac{1}{5}\sqrt{\frac{4}{\pi}\frac{L_x\cdot L_z}{n_\mathrm{seeds}}}$).

An arbitrary angle between 0 $^{\circ}$ and 60 $^{\circ}$ has been assigned to each nucleation site to define the orientation of the crystalline growth of graphene.
Under-coordinated atoms on the edge of each grain have been added to a list of possible reactive sites. In an iterative process, the reactive site to which the next atom has been added, has been randomly selected from this list.
The list of reactive atoms has been updated in each step until no reactive atoms were left.
Grain growth has been terminated, i e. atoms have been eliminated from the list of reactive atoms, when the respective C atom was three-fold coordinated or atoms from neighboring cells were closer than 95\% of the equilibrium C-C bond length (here: 1.3978 \AA).


Creation of simulation cells by this method only considers geometrical aspects. Interatomic interactions are neglected. Furthermore, simulation cells are created only in two dimensions, omitting corrugation in off-plane direction. To account for the latter, simulation cells have been relaxed during molecular dynamics (MD) simulations at high temperatures. Details on the MD simulations can be found in Section~\ref{sec:aemd}. Velocity rescaling has been applied for this equilibration process using a time step of 0.5 fs. It has been initiated at a temperature of 300 K for 10 ps. Subsequently, the cells have been heated to a temperature of 4000 K with a heating rate of 105 K/ps (in 35 ps) and relaxed at this temperature for 255 ps. The cells have been cooled back to 300 K with a cooling rate of  10 K/ps (in 370 ps). This procedure has been followed by a microcanonical (NVE) run for 200 ps.

Several geometrical criteria have been used to characterize the grain size of the created nanocrystalline samples. One possible way is the determination of an ideal radius $r_\mathrm{id}$ assuming the particles to have a circular shape calculated from the area of the grain which is estimated from the number of atoms in the grain ($r_\mathrm{id}=\sqrt{\frac{A_\mathrm{G}}{\pi}}$).
However, this characterization does not consider different shapes of the grains. Therefore, we also included the characterization by the radius of gyration $r_{\mathrm{G},j}$, which has been determined for each seed $j$ according to Eq.~\ref{eq:rgyr}, where $\vec{r}_{i,j}$ is the position of atom $i$ in grain $j$ and $r_{CM,j}$ and $n_{\mathrm{C},j}$ are the center of mass and the number of atoms in the grain, respectively.

\begin{equation} \label{eq:rgyr}
r_{\mathrm{G},j}^2=\frac{\sum_i^{n_{\mathrm{C},j}}\left(\vec{r}_{i,j}-\vec{r}_{CM,j}\right)^2}{n_{\mathrm{C},j}}
\end{equation}

At least two different configurations have been created for each sample dimension and grain size. For systems with an average radius of gyration of 1 nm, the transverse section $L_x$ orthogonal to the direction of the heat flow has been changed from 5 to 20 nm. Furthermore, for such systems, the sample length $L_z$ in direction of the thermal transport has been changed from 50 to 1000 nm. The effect of the sample length has been investigated additionally for samples with $r_\mathrm{G}$=2.5 nm and $L_x$=20 nm. These calculations have been carried out to verify the reliability of the used simulation cells for the determination of the thermal conductivity. Based on these results, the average radius of gyration has been changed from 0.7 nm to 14 nm using samples with a cell length of 200 nm.

\begin{figure}[tb]
\centering
  \includegraphics[width=0.4\textwidth]{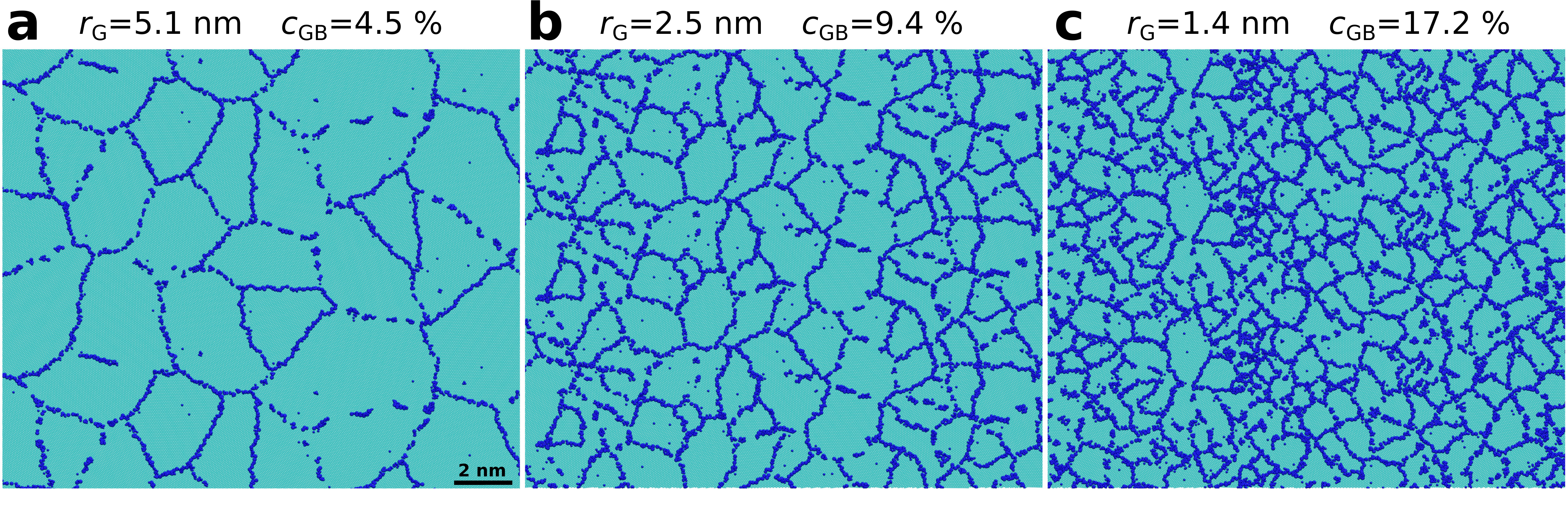}
       \caption{Simulation cells of nanocrystalline graphene with a mean radius of gyration of (a) 5.1, (b) 2.5 and (c) 1.4 nm. Atoms assigned to the grain boundary are indicated in blue.}
       \label{fig:cell}
\end{figure}

Atoms belonging to the grain boundaries have been identified based on their energy (energy/atom). For this purpose, the energy/atom distribution has been evaluated for initial configurations of each sample. It shows a main peak for the atoms belonging to the center of the grains with a shoulder evolving at higher energies corresponding to atoms in the grain boundaries. Atoms with energies higher than 2\% of the main peak value have been assigned to the grain boundaries. By this technique some atoms from the center of the grain might be selected as a result of corrugation in out-of-plane direction. However, the amount is negligible. Nanocrystalline graphene grain boundary atoms selected in this manner are shown in Figure~\ref{fig:cell} for samples with grain sizes of $r_\mathrm{G}$=1.4, 2.5 and 5.1 nm. This characterization allowed us to determine the concentration of grain boundary atoms $c_\mathrm{GB}$ in each sample.

\subsection{AEMD simulations}
\label{sec:aemd}
Molecular dynamics simulations have been performed using the \textsc{lammps} code.\cite{Plimpton1995,Plimptop} Covalent interactions between carbon atoms have been described by the second-generation reactive empirical bond order (REBO) potential which has previously been shown to give reasonable results for graphene-based systems.\cite{Barbarino2014,Barbarino2015a}

The thermal conductivity has been determined based on the AEMD methodology. In this formalism, the simulation cell is firstly divided into two regions with equal length $L_z$/2. One of these two compartments is equilibrated at a high temperature ($T_\mathrm{h}$=400 K), the other compartment at a low temperature ($T_\mathrm{c}$=200 K) by velocity rescaling. This creates an initial step-like function of the temperature along the sample length in $z$-direction.\cite{Melis2014a} Next, the evolution of the average temperature in the hot ($T_\mathrm{h}$) and cold ($T_\mathrm{c}$) reservoir is recorded during a transient regime towards equilibrium of microcanonical evolution. Based on Fourier's theorem of thermal transport and the given step-like initial temperature profile, the evolution of the temperature gradient ($\Delta T = T_\mathrm{h}-T_\mathrm{c}$) follows 
$\Delta T\left( t \right)=\sum_{n=1}^\infty C_n e^{\alpha_n^2\bar{\kappa} t},$
where $\bar{\kappa}=\frac{\kappa}{\rho c_v}$ is the thermal diffusivity with the density $\rho$ of the material and its specific heat $c_v$.
This expression is fitted to the temperature gradient obtained from the simulations to determine $\kappa$. More details on the AEMD methodology can be found elsewhere.\cite{Melis2014a,Lampin2013}

\section{Results and Discussion}
\subsection{Morphology of nanocrystalline graphene samples}
\begin{figure}[tb]
\centering
  \includegraphics[width=0.4\textwidth]{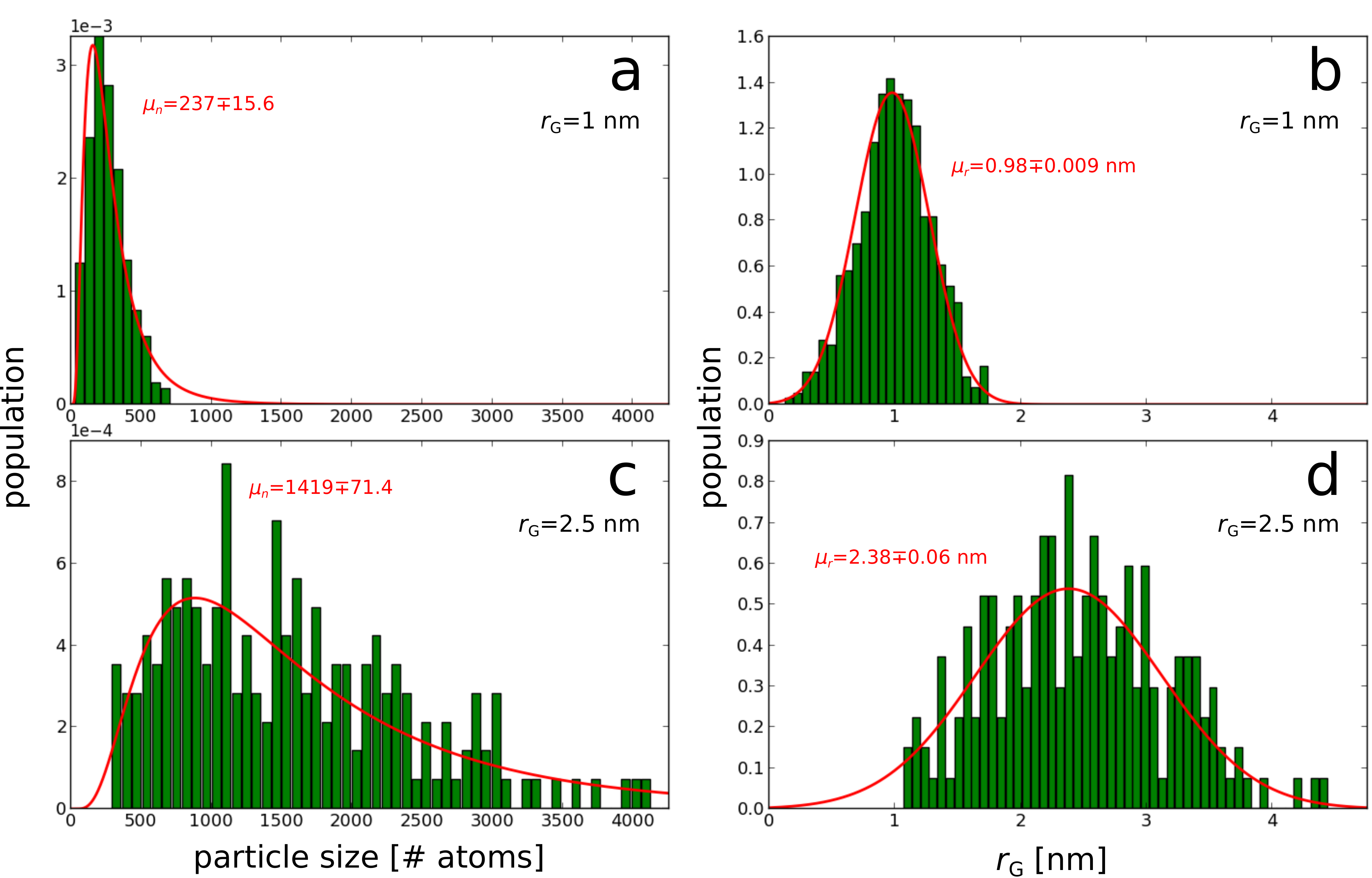}
       \caption{PSD of nanocrystalline graphene with an average radius of gyration $r_\mathrm{G}$ of (a,b) 1 and (c,d) 2.5 nm. The PSD follows a logarithmic normal distribution as a function of the number of atoms per grain (a,c) and a normal distribution as a function of $r_\mathrm{G}$ (b,d).}
       \label{fig:log_norm_var}
\end{figure}

Nanocrystalline graphene samples have been characterized by the number of C atoms $n_{\mathrm{C}}$ in each grain and by their radius of gyration $r_\mathrm{G}$ which has been determined according to Eq.~\ref{eq:rgyr}. The particle size distribution (PSD) based on $n_\mathrm{C}$ is best described by a lognormal distribution as shown in Figure~\ref{fig:log_norm_var} on the example of samples with average $r_\mathrm{G}$ of 1 and 2.5 nm. The data has been fitted to

\begin{equation}
\label{eq:log}
f(x)=\frac{1}{x\sqrt{2\pi}\sigma}e^{-\frac{\left(\ln{x}-\tilde{\mu}_n\right)^2}{2\sigma^2}}
\end{equation}

resulting in a mean number of atoms per grain ($\mu_n=e^{\tilde{\mu}_n}$) of 237 and 1419, respectively. 
The PSD based on the radius of gyration, however, is better described by a normal distribution and accordingly has been fitted to

\begin{equation}
\label{eq:norm_grain}
f(x)=\frac{1}{\sqrt{2\pi}\sigma}e^{-\frac{\left(x-\mu_r\right)^2}{2\sigma^2}}.
\end{equation}

\begin{figure}[tb]
\centering
  \includegraphics[width=0.4\textwidth]{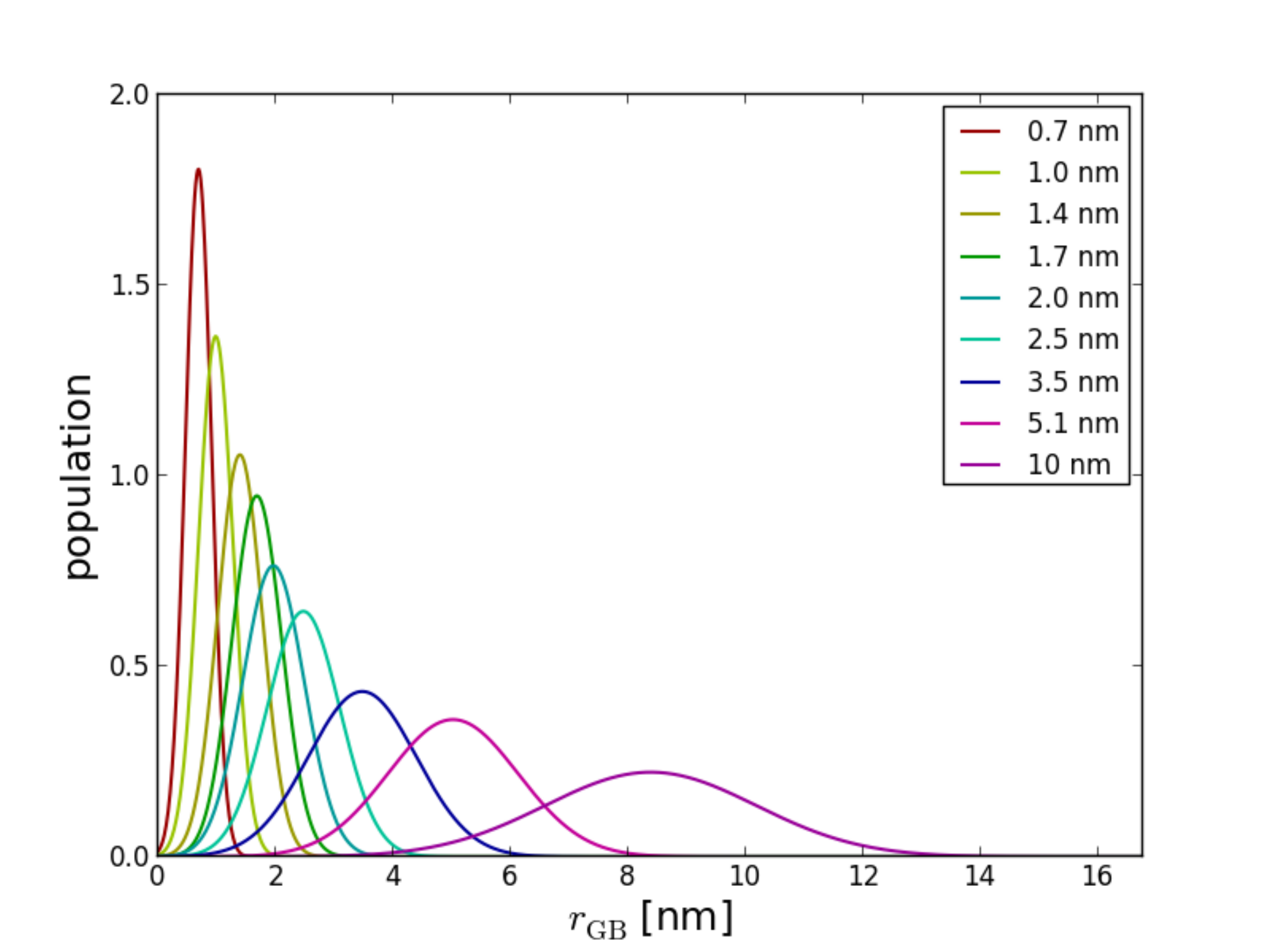}
       \caption{Normal distribution of nanocrystalline samples with average radii of gyration from 0.7 to 10 nm.}
       \label{fig:psd_all}
\end{figure}

For a sufficiently large number of grains, the mean radius of gyration $\mu_r$ obtained from Eq.~\ref{eq:norm_grain} is equivalent to its arithmetic mean. Small variations, however, have been observed when the number of seeds is respectively low and not sufficient for a statistical analysis. In a simulation cell with dimensions 40 nm x 200 nm and a total number of 200 grains ($r_\mathrm{G}=$ 2.5 nm, Figure~\ref{fig:log_norm_var}), distributed in 50 bins, the mean of the normal distribution is slightly lower (2.38 nm) than the arithmetic mean.
The fitted normal distribution of simulated samples with $L_z$=200 nm and varied grain sizes from 0.7 to 10 nm are shown in Figure~\ref{fig:psd_all}.

\begin{figure}[tb]
\centering
   \includegraphics[width=0.4\textwidth]{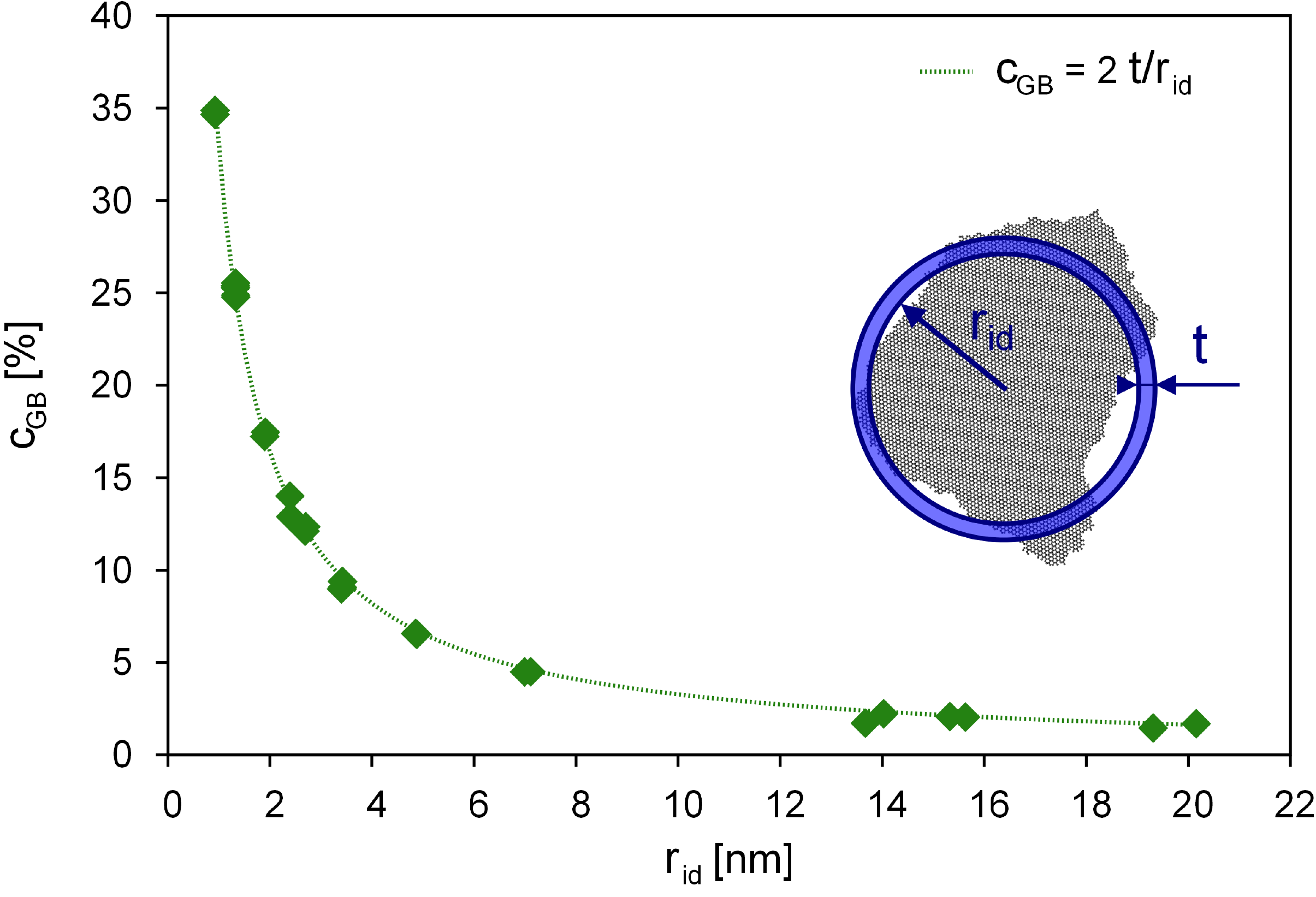}
    \caption{Number of atoms in the boundary region as a function of the radius $r_{\mathrm{id}}$, assuming ideal circular shape.}
    \label{fig:c_rid}
\end{figure}

Grain boundaries have been further analyzed based on the number (concentration) of atoms in the grain boundary $c_\mathrm{GB}$. Assuming ideal circular shape of the grains, $c_\mathrm{GB}$ is related to the particle radius $r_\mathrm{id}$ according to
\begin{equation}
\label{eq:c_gb}
c_\mathrm{GB}=\frac{A_\mathrm{GB}}{A_\mathrm{G}}=2\frac{t}{r_\mathrm{id}}
\end{equation}

where $A_\mathrm{G}$ is the total the area of one grain and $A_\mathrm{GB}$ is the area of the grain boundary with a thickness $t$. Fitting $c_\mathrm{GB}$ to eq. \eqref{eq:c_gb} results in a thickness of 0.164 nm (Figure~\ref{fig:c_rid}). 
From the density (atoms/area) of various samples calculate here we estimated the spacial extension of one atomic layer to be 0.16 nm. We therefore conclude that the grain boundaries in nanocrystalline graphene consist in average of one atomic layer of C atoms.

\subsection{Sample size effects}
\subsubsection{Length-dependent thermal conductivity}
The effect of the sample length $L_z$ on the estimated thermal conductivity has been investigated for samples with average grain size $r_\mathrm{G}$ of 1 and 2.5 nm. The sample length has been increased from 50 nm to 1 $\upmu$m.
The size effect on thermal conductivity of crystalline graphene is under major dispute in recent literature.
Several experimental and theoretical studies have claimed divergence of the thermal conductivity with the length of the sample.\cite{Xu2014,Klemens1994,Klemens2001} However, more recently it has been proven that, while thermal conductivity grows logarithmically with the sample length up to $\sim$10 $\upmu$m, it is definitely upper limited in samples long enough to allow a diffusion transport regime for both single and collective phonon excitations.\cite{Seol2010,Fugallo2014,Barbarino2015a}

Accordingly, here, we describe the length dependent thermal conductivity of crystalline and nanocrystalline  graphene with the linear reciprocal (rational) function 
\begin{equation} \label{eq:k_lin}
\begin{split}
\kappa &= \frac{\kappa_\infty}{\left(1+\frac{\lambda}{L_z}\right)} \\
\frac{1}{\kappa} &= \frac{1}{\kappa_\infty}\left(1+\frac{\lambda}{L_z}\right)
\end{split}
\end{equation}
from which the convergent bulk thermal conductivity $\kappa_\infty$ at infinite sample size can be estimated.

\begin{figure}[htb]
\centering
  \includegraphics[width=0.4\textwidth]{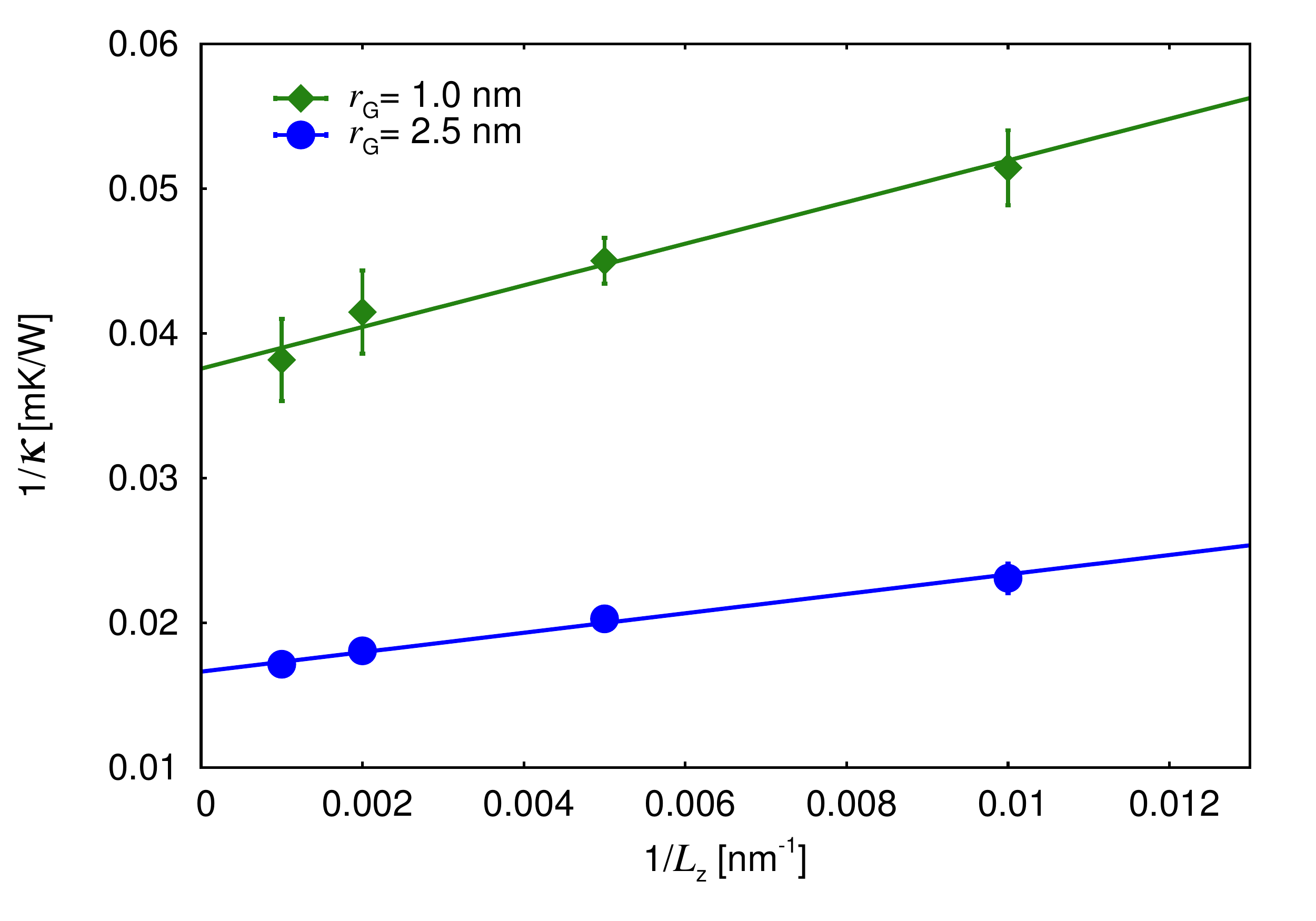}
       \caption{Inverse of thermal conductivity 1/$\kappa$ as a function of the sample length 1/$L_z$ of nanocrystalline graphene with average grain size $r_\mathrm{G}=$ 1 and 2.5 nm.}
       \label{fig:K_L_Rvar}
\end{figure}

In the studied range of sample sizes both in crystalline and the two nanocystalline samples, the 1/$\kappa$ to 1/$L_z$ behavior follows a linear trend according to Eq.~\ref{eq:k_lin} (Figure~\ref{fig:K_L_Rvar}) indicating that phonon properties in such materials can be approximated well by an average value.\cite{Sellan2010}
A bulk thermal conductivity $\kappa_\infty$ of 60.2 and 26.6 W/mK (Table~\ref{tab:k_vs_l})  has been estimated for nanocrystalline graphene with $r_\mathrm{G}$ of 2.5 and 1 nm, respectively. These values correspond to 7 and 3\%, respectively, of the bulk thermal conductivity of crystalline graphene (859.7 W/mK). The relative change of the thermal conductivity with respect to the one of pristine graphene is in excellent agreement with a recent study where the thermal conductivity of polycrystalline graphene has been estimated using equilibrium molecular dynamics simulations.\cite{Mortazavi2014}

\begin{table} [bt]
\caption{Optimized parameters $\kappa_\infty$ and $\lambda$ of the approximation function Eq.~\ref{eq:k_lin} for the thermal conductivity of nanocrystalline graphene with average grain size of 1 and 2.5 nm and of pristine graphene as a function of the sample size.}
\label{tab:k_vs_l}
\begin{center}
\begin{tabular}{cccc}
\hline
\rule{0pt}{3ex}$r_\mathrm{G}$ [nm] & $\kappa_\infty$ $\left[\frac{\mathrm{W}}{\mathrm{mK}}\right]$ & $\lambda$ [nm] & $L_x$ [nm]\rule{0pt}{3ex}\\
\hline
  \rule{0pt}{3ex}1 &  25.2$\pm$0.3 & 27.7$\pm$3.0 & 5\\
  1 &  23.9$\pm$0.1 & 22.4$\pm$1.1 & 10\\
  1 &  26.6$\pm$0.6 & 38.3$\pm$5.8 & 5-80\\
  2.5 & 60.2$\pm$0.7 & 40.4$\pm$3.3& 5-80\\
  crystalline & 859.7$\pm$29.6 &  455.9$\pm$40.0 &3-10\\
\hline
\end{tabular}
\end{center}
\end{table}

The characteristic length $\lambda$, on the other hand, changes only marginally with the grain size. It has been estimated to be 40.4 and 38.3 nm for grains with $r_\mathrm{G}$ of 2.5 and 1 nm, respectively.
This indicates similar convergence behavior of the thermal conductivity in the two samples.
In fact, according to Eq.~\ref{eq:k_lin}, 90\% of the bulk thermal conductivity are obtained at a sample length $L_z$ of 364 nm and 345 nm for a grain size of 2.5 nm and 1 nm, respectively. In contrast, in crystalline graphene, samples with a length of more than 4 $\upmu$m ($\lambda=455.9$ nm) have to be simulated to reach 90\% of the bulk thermal conductivity.

These results demonstrate that the mean free path (MFP) of dominant phonons present in crystalline graphene is drastically reduced when grain boundaries in the form of nanosized grains are introduced. The size of the grains, however, influences the MFP of dominant phonons only marginally, at least for the grain sizes studied here (2.5 and 1 nm).

\subsubsection{Effect of transverse section}
The effect of the transverse of the samples on the estimated thermal conductivity has been evaluated for samples with average grain size of $r_\mathrm{G}$=1 nm where an intersection of 5 and 10 nm has been simulated. Several configurations have been considered at each length.

For an intersection of 5 nm,  $\kappa_\infty$ results in 25.2 W/mK and the characteristic length $\lambda$ is estimated with 27.7 nm. 
Increase of the intersection had only little effect on the thermal conductivity. At an intersection of 10 nm, $\kappa_\infty$ and $\lambda$ have been determined to be 23.9 W/mK and 22.4 nm, respectively (Table~\ref{tab:k_vs_l}). We thus consider an intersection of 5 nm to be sufficiently large for a reasonable estimation of $\kappa$.


\subsection{Accumulation of thermal conductivity}
The spectral contribution of phonons to the thermal conductivity has been estimated by an accumulation function. The accumulation function describes the ratio of the estimated thermal conductivity at a certain sample length to the bulk thermal conductivity.\cite{Sellan2010,Hahn2014} It can thus be understood as the contribution to the thermal conductivity of phonons that have a MFP ($\Lambda$) smaller than the corresponding sample length $L_z$ according to 

\begin{equation}
\label{eq:accum}
\frac{\kappa\left(\mathrm{\Lambda}\right)}{\kappa_\infty}=\int_0^\mathrm{\Lambda} f\left(\tilde{\mathrm{\Lambda}}\right)d\tilde{\mathrm{\Lambda}},
\end{equation}
where $f\left(\mathrm{\Lambda}\right)$ is the distribution function of the MFP of dominating phonons.

A normal distribution has been assumed for $f\left(\mathrm{\Lambda}\right)$, where
$\mathrm{\Lambda}$ enters into the equation as the logarithm of $L_z$ and has been normalized to the characteristic length $\lambda$ ($\mathrm{\Lambda}=\log\left(\frac{L_z}{\lambda}\right)$). Following Ref. [\cite{Hahn2014}] we set

\begin{equation}
\label{eq:norm}
f\left(\mathrm{\Lambda}\right)=\frac{1}{\sigma \sqrt{2\pi}}\exp\left(-\frac{\left(\mathrm{\Lambda}-\tilde{\mu}\right)^2}{2\sigma^2}\right)
\end{equation}

The parameters $\tilde{\mu}$ and $\sigma$ of the normal distribution of the phonon MFP have been estimated fitting the accumulation of thermal conductivity as a function of the sample lengths $L_z$ to Eq.~\ref{eq:erf}. The statistical value $\tilde{\mu}$ can be translated into $\mu_\mathrm{MFP}=\lambda \cdot 10^{\tilde{\mu}}$ which corresponds to the average value of the phononic MFP of the system.

\begin{equation}
\label{eq:erf}
\frac{\kappa\left( \mathrm{\Lambda}\right)}{\kappa_\infty}=\frac{1}{2}\left[1+\mathrm{erf}\left(\frac{\mathrm{\Lambda}-\tilde{\mu}}{\sigma\sqrt{2}}\right)\right]
\end{equation}

\begin{figure}[tb]
\centering
 \includegraphics[width=0.4\textwidth]{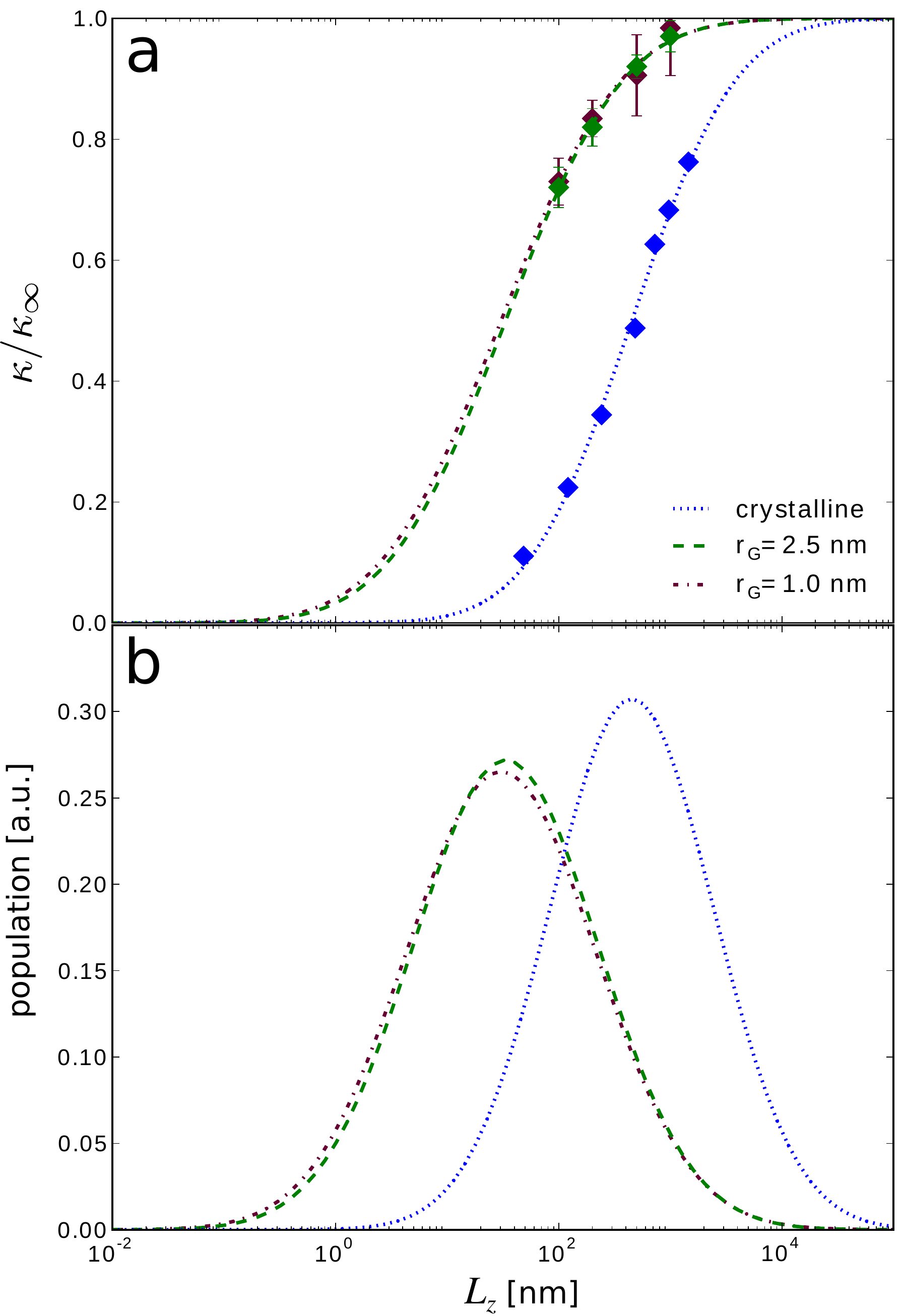}
      \caption{(a) Accumulation and (b) distribution function of nanocrystalline and crystalline graphene.}
      \label{fig:accum_all}
\end{figure}

Figure~\ref{fig:accum_all}a shows the accumulation function that has been optimized for both nanocrystalline systems in comparison to the results of crystalline graphene. The corresponding normal distribution is represented in Figure~\ref{fig:accum_all}b along with the optimized parameters in Table~\ref{tab:norm_dist}.

\begin{table} [bt]
\caption{Estimated values of variance $\sigma^2$ and average mean free path $\mu_\mathrm{MFP}$ of dominant phonons.}
\label{tab:norm_dist}
\begin{center}
\begin{tabular}{cccc}
\hline
\rule{0pt}{3ex}$r_\mathrm{G}$ [nm] & $\mu_\mathrm{MFP}$ [nm]& $\sigma^2$ & $R^2$\rule{0pt}{3ex}\\
\hline
  \rule{0pt}{3ex}1 & 30.46 $\pm$ 0.107 & 0.718 & 0.978\\
  2.5 & 33.46 $\pm$ 0.036 & 0.686 & 0.997\\
  crystalline & 451.38 $\pm$  0.015 & 0.535 & 0.995\\
\hline
\end{tabular}
\end{center}
\end{table}

The average MFP of dominant phonons $\mu_\mathrm{MFP}$ in nanocrystalline graphene with a grain size of 2.5 and 1 nm is estimated to be 30.5 and 33.5 nm, respectively, indicating that decrease of the grain size from 2.5 to 1 nm hardly changes the MFP spectrum of dominant phonons.
In both cases, a significant effect of the grain size on the bulk thermal conductivity is observed.
We therefore argue that lattice vibrations in the nanocrystalline material are similar to the vibrations in pristine graphene but with additional contributions from phonon scattering at the grain boundaries.

\subsection{Analysis of the vibrational density of states}
In order to better characterize the phonon spectrum in nanocrystalline graphene, the vibrational density of states (VDOS) has been estimated for several samples including crystalline graphene and nanocrystalline graphene with $r_\mathrm{G}$=0.7, 1, 2.5 and 5.1 nm. Furthermore, the VDOS has been calculated specifically of atoms in the grain boundary.
The VDOS has been calculated as the Fourier transform of the velocity autocorrelation function where the atomic velocities have been collected during a canonical simulation (NVT) at 300 K for 10 ps preceded by a canonical equilibration at 300 K for 200 ps. The accuracy of this method has been compared to results obtained from density functional theory (DFT) calculations through the diagonalization of the dynamical matrix\cite{Cadelano2010} and have been shown to be in general agreement.
\begin{figure}[tb]
\centering
   \includegraphics[width=0.4\textwidth]{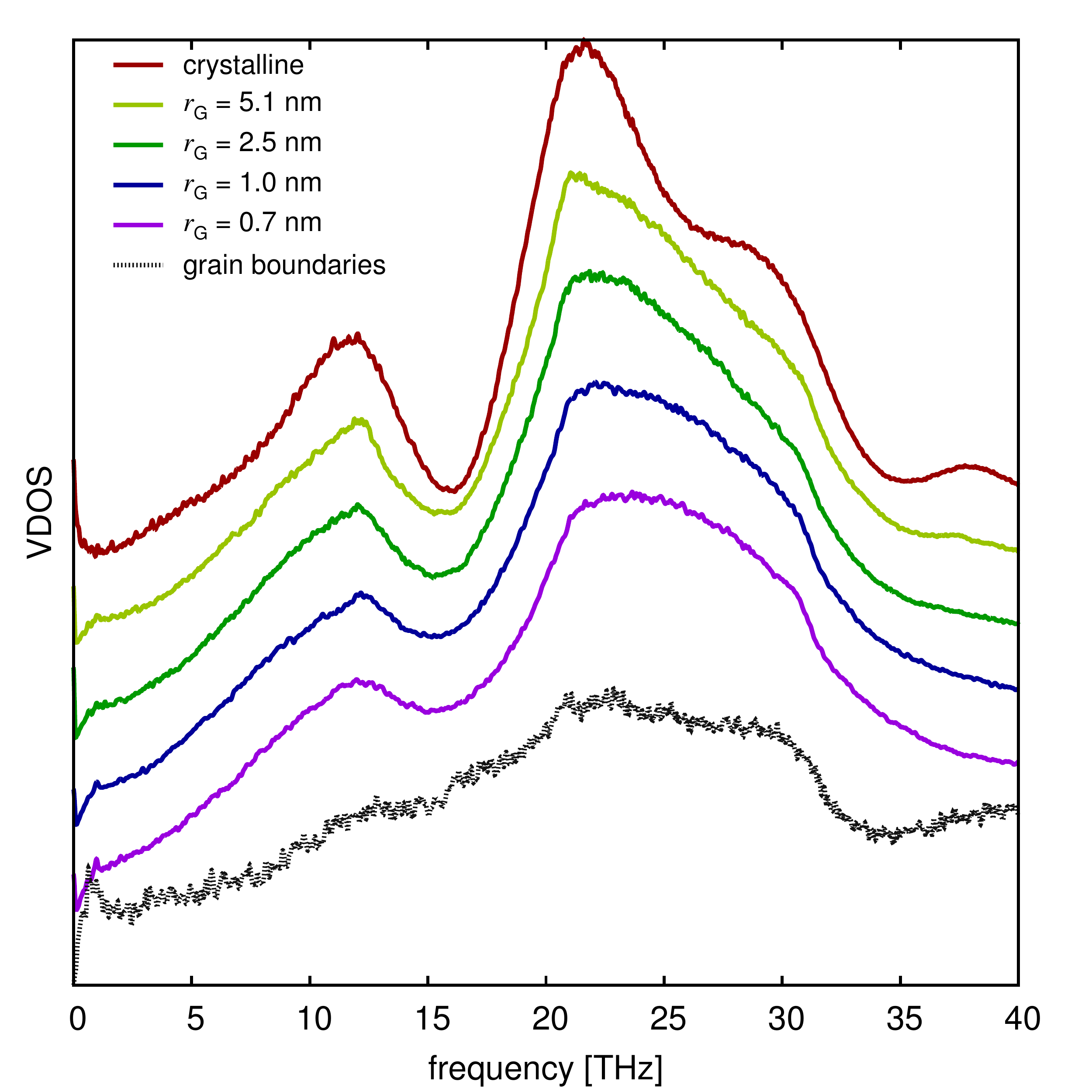}
    \caption{Vibrational density of states (VDOS) of crystalline and nanocrystalline graphene in the range of acoustic phonons.}
    \label{fig:vdos}
\end{figure}
This is in particular the case in the acoustic range of phonon modes ($<$50THz) which are mainly contributing to thermal transport in graphene verifying the reliability of our model to describe the VDOS in graphene-based materials. 


Figure~\ref{fig:vdos} shows the VDOS of acoustic phonons of pristine and nanocrystalline graphene and of the atoms belonging to the grain boundary. The main peaks of acoustic phonons in crystalline graphene are observed at $\sim$12 and $\sim$22 THz. The peak at 22 THz is extended by a shoulder to higher frequencies resulting from an additional peak at $\sim$28 THz. In nanocrystalline graphene, an overall broadening of all peaks is observed combined with a decrease in intensity. In particular, a shoulder at lower frequencies of the 12 THz peak seems to evolve which indicates enhanced vibrations at $\sim$9 THz. Similarly, the shoulder at higher energies of the 22 THz peak which is most pronounced at $\sim$28 THz flattens out in nanocrystalline samples which can possibly result from additional vibrations at $\sim$25 THz. These observations are confirmed regarding the VDOS of grain boundary atoms (black line). There, the main vibrations of crystalline graphene at 12 and 22 THz disappeared due to the absence of any ordered crystalline structure. Instead, the density of vibrations increases monotonously from 0 to 22 THz and is basically unchanged for frequencies from 22 to 30 THz. This indicates enhanced vibrational states between 15 and 20 THz and between 25 and 28 THz with respect to crystalline graphene. In summary, we consider the decrease in intensity and the broadening of the peaks as the main reasons for the observed decreased in the thermal conductivity of nanocrystalline graphene. Whereas phonon scattering at the grain boundaries can be assigned to vibrations between 15 and 20 THz and between 25 and 28 THz.

\subsection{Grain size effects}
In the interest of computational costs, the effect of the grain size in nanocrystalline graphene on its thermal conductivity has been studied at a finite sample length of $L_z$=200 nm. 

For an intersection of 20 nm, samples with initial seed densities from 0.025 to 0.32 seeds/nm$^2$, resulting in grain sizes  $r_\mathrm{G}$ from 0.7 to 2.5 nm, have been simulated. To determine the thermal conductivity of larger grains, the intersection had to be increased to 40 nm and 80 nm. Initial seed densities have been reduced down to 7.5$\times 10^{-4}$ seeds/nm$^2$ leading grains sizes up to 14 nm.
Following the results of the previous paragraphs, marginal deviation in the estimated thermal conductivity is expected when the intersectional length $L_x$ is varied. Thus, results of the thermal conductivity of samples with different $L_x$ will be compared directly.

\begin{figure}[htb]
\centering
   \includegraphics[width=0.4\textwidth]{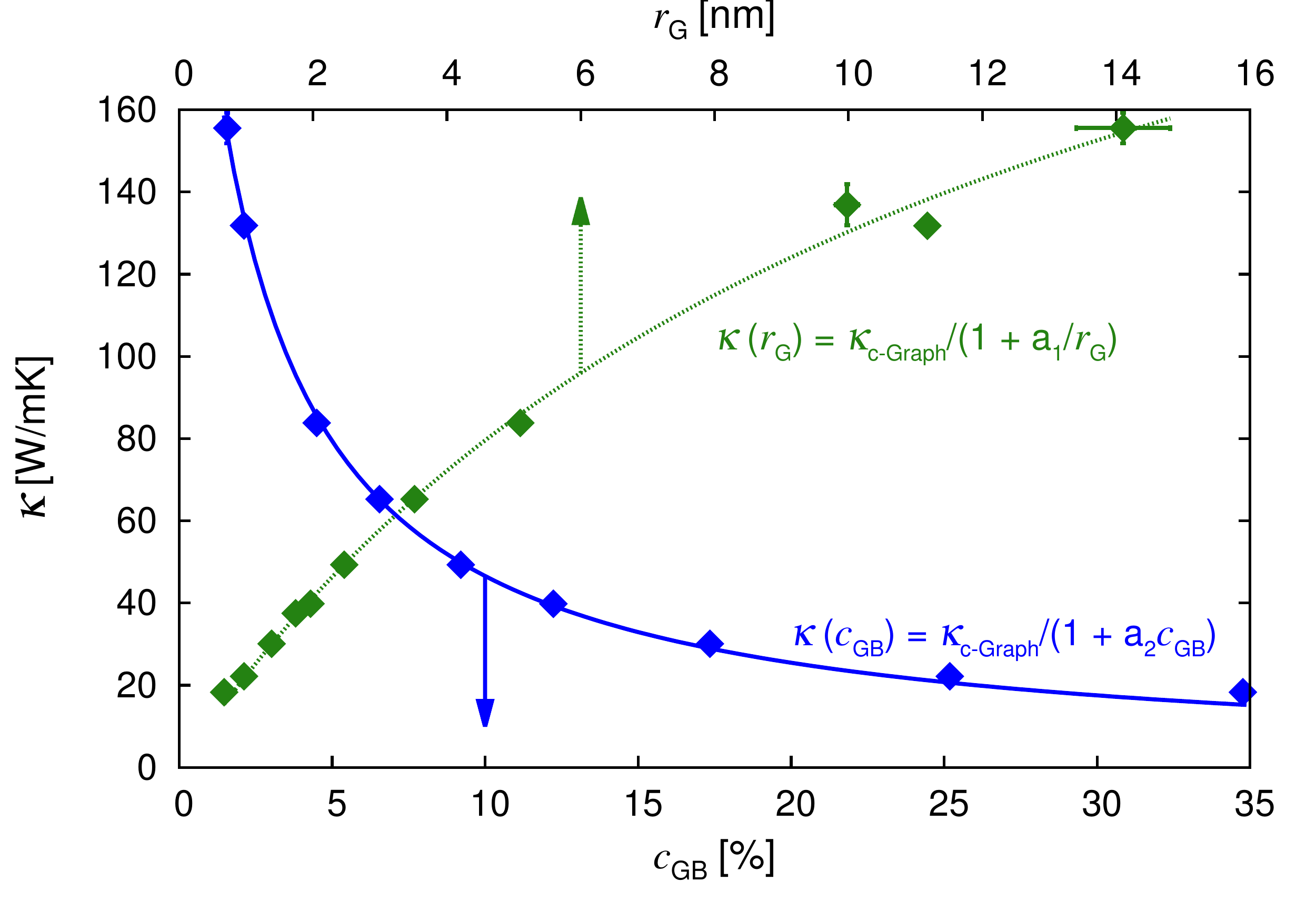}
    \caption{Thermal conductivity of nanocrystalline graphene $\kappa$ as a function of the concentration of atoms in the grain boundary $c_\mathrm{GB}$ (solid line) and of the radius of gyration $r_\mathrm{G}$ (dotted line).}
    \label{fig:k_vs_n}
\end{figure}

Thermal conductivity of nanocrystalline graphene as a function of the grain size is shown in Figure~\ref{fig:k_vs_n}. It increases continuously from 18.3 to 156 W/mK for increasing grain size ($r_\mathrm{G}$) from 0.7 to 14 nm. The introduction of grain boundaries in the crystalline graphene can be regarded as a connection in series of resistances. Based on this a rational function is used to describe the dependency of $\kappa$ on $r_\mathrm{G}$ (Eq.~\ref{eq:k_vs_r}) where $\kappa_\mathrm{c-Graph}$ is the thermal conductivity of crystalline grains which is assumed to be constant.

\begin{equation} \label{eq:k_vs_r}
\begin{split}
\frac{1}{\kappa}\left(r_\mathrm{G}\right) &= \frac{1}{\kappa_\mathrm{c-Graph}}+\frac{R_\mathrm{GB}}{2r_\mathrm{G}} \\
\kappa\left(r_\mathrm{G}\right) &= \frac{\kappa_\mathrm{c-Graph}}{1+\frac{R_\mathrm{GB}\kappa_\mathrm{c-Graph}}{2r_\mathrm{G}}}
\end{split}
\end{equation}

A similar argumentation has been used previously where the thermal conductivity of polycrystalline graphene has been calculated using equilibrium molecular dynamics.\cite{Mortazavi2014}

The parameters $\kappa_\mathrm{c-Graph}$ and $R_\mathrm{GB}$ of Eq.~\ref{eq:k_vs_r} have been optimized with 282 W/mK and 8.2$\times 10^{-11}$ m$^2$K/W, respectively.
It has to be noted that the value of the thermal resistance might not be representative since it is very sensitive to the value of $r_\mathrm{G}$ which is only a rough estimation of geometrical properties of the grains. It gives indications for the thermal resistance but should rather be regarded as a fitting parameter for the description of $\kappa$ as a function of $r_\mathrm{G}$.
The constructed constant thermal conductivity $\kappa_\mathrm{c-Graph}$ slightly overestimates the thermal conductivity of pristine graphene at this cell length (262 W/mK) by 7.5 \%.
$\kappa_\mathrm{c-Graph}$ thus overestimates the actual value of crystalline graphene at this cell length by 24\%.

This is striking since it predicts that the thermal conductivity of the crystalline domains (with a length smaller than the simulation cell) would be higher than the reference value of the thermal conductivity of crystalline graphene in a simulation cell with the same dimensions. However, it can partly be explained by the fundamentals of the approximation function of $\kappa\left(r_\mathrm{G}\right)$.
In Eq.~\ref{eq:k_vs_r}, the number of grain boundaries contributing to the thermal resistance $R_\mathrm{GB}$ is assumed to be the same as the number of grains $n_\mathrm{seeds}$, instead it should be reduced by one ($n_\mathrm{GB}=n_\mathrm{seeds}-1$). This approximation is valid for a reasonably large number of grains ($\frac{n_\mathrm{seeds}-1}{n_\mathrm{seeds}}\rightarrow 1$). However, since simulations have been carried out at a finite sample length $L_z$, $n_\mathrm{seeds}$ decreases with increasing $r_\mathrm{G}$.
Thus Eq.~\ref{eq:k_vs_r} overestimates the actual thermal conductivity for large $r_\mathrm{G}$.

In Eq.~\ref{eq:k_vs_r} it is assumed that crystalline domains are spherical with a radius $r_\mathrm{G}$ and the length of the boundaries perpendicular to thermal transport, effectively contributing to the thermal resistance $R_\mathrm{GB}$, is equal to the diameter (2$r_\mathrm{GB}$) of the grains.
To account for the actual geometry of grains, the thermal conductivity has been described as a function of the number (concentration) of atoms in the grain boundaries ($c_\mathrm{GB}$) according to

\begin{equation} \label{eq:k_vs_c}
\begin{split}
\kappa\left(c_\mathrm{GB}\right) &= \frac{\kappa_\mathrm{c-Graph}}{1+\frac{\kappa_\mathrm{c-Graph}\cdot R_\mathrm{GB}}{d_{l}} c_\mathrm{GB}}
\end{split}
\end{equation}

where $d_{l}$ corresponds to the thickness of one atomic layer of C atoms. For the samples calculated here an average value of 0.16 nm has been used.
Applying this expression for the overall thermal conductivity $\kappa$ of the nanocrystalline simulation cell, the thermal boundary resistance $R_\mathrm{GB}$ is approximated with 2.8 $\times 10^{-11}$ m$^2$K/W and the constant thermal conductivity of the crystalline domains $\kappa_\mathrm{c-Graph}$ is found to be 269 W/mK. As mentioned before, $R_\mathrm{GB}$ will be regarded only as a fitting parameter rather than the actual physical value of the thermal resistance.
The constant thermal conductivity $\kappa_\mathrm{c-Graph}$, on the other hand, can directly be compared to the thermal conductivity of crystalline graphene. It still overestimates the value of crystalline graphene (262 W/mK), however, in this case only by 2.6\%. This demonstrates the relevance to account for the actual properties of the grain boundaries (density of such) when estimating the thermal transport.

Both approximations indicate that the thermal conductivity of the crystalline domain $\kappa_\mathrm{c-Graph}$ is estimated well by its bulk-like value. It demonstrates that the thermal conductivity of nanostructured systems such as polycrystalline graphene can be estimated considering grain boundaries and crystalline parts as a connection of resistances in series. In this scheme phonon scattering effects are included in the thermal resistance $R_\mathrm{GB}$ of the grain boundaries and the thermal conductivity of crystalline domains can be described by the thermal conductivity of the crystal without grain boundaries. However, here, grain boundaries are relatively small and can be described as infinitely small. It is supposed that for systems with larger grain boundaries, phonon scattering effects cannot only be included in the thermal boundary resistance but need to be included as a separate length dependent resistance/conductivity.

\section{Conclusions}
An iterative algorithm has been implemented to create samples of 2D nanocrystalline graphene sheets respecting periodic boundary conditions. The average grain size of single-crystalline domains has been controlled by the initial density of seeds per unit area. The grain size has been characterized by the radius of gyration, the number of particles per atom and the concentration of atoms belonging to the grain boundaries. The grain size distribution of the nanocrystalline materials described by the radius of gyration follows normal distribution. Characterizing the size of particles by their area, i.e. number of particles per grain, a lognormal distribution of the particle size has been found. The thickness of grain boundaries has been estimated from the change in concentration of grain boundary atoms as a function of the idealized circular radius. It has been estimated to be 0.16 nm corresponding to one atomic layer.

Thermal transport of such nanocrystalline graphene sheets has been exploited using AEMD simulations at an average temperature of 300 K. The sample size dependence of the thermal conductivity has been investigated in nanocrystalline samples with an average grain size of 1 and 2.5 nm showing a linear trend between 1/$\kappa$ and 1/$L_z$. Based on this, the bulk thermal conductivity at infinite sample length could be estimated.
In a sample with average grain size of 1 nm the bulk thermal conductivity has been estimated to be 26.6 W/mK which corresponds to 3\% of the thermal conductivity of crystalline graphene.

Based on the accumulation function of the thermal conductivity, the average mean free path of phonons contributing to thermal conductivity has been estimated for both nanocrystalline samples and crystalline graphene. Introduction of grain boundaries resulted in a remarkable decrease of the estimated mean free path of crystalline graphene (451 nm) to ~30 nm. The grain size, however, had only marginal effect on the average mean free path.

Analysis of the vibrational density of states (VDOS) of crystalline and nanocrystalline graphene showed a general decrease in the density of vibrational states and a broadening of the peaks when grain boundaries in form of nanograins are introduced. The VDOS of grain boundary atoms in particular revealed increased vibrations between 15 and 20 Hz and between 25 and 28 Hz which can be assigned to phonon scattering in the boundary layer and lead to the observed broadening of the VDOS in nanocrystalline graphene.

Thermal conductivity at a sample length of 200 nm has been analysed as a function of the grain size. In the investigated range ($r_\mathrm{G}$ = 0.7 to 14 nm), the thermal conductivity can be approximated well with a rational function based on the idea of grain boundaries to behave as thermal resistances connected in series.

A better description of the overall thermal conductivity of the nanostructured system, however, is obtained when the concentration of grain boundary atoms is considered. It is proposed that for such nanostructured systems, the overall thermal resistance can be described well as a connection of series where phonon scattering effects are included solely in the resistance of the grain boundary and the conductivity of the crystalline domains can be estimated by the conductivity of a crystal without grain boundaries.

\begin{acknowledgments}
This work is financially supported by the SNF grant with the project number P2ZHP2\_148667. Simulations have been conducted on the HPC resources of CINECA under the project ISCRA\_THETRASI. We also acknowledge financial support by MIUR under project PRIN 2010-2011 GRAF. CM additionally acknowledges Sardinian Regional Government for financial support (P.O.R. Sardegna ESF 2007-13).
\end{acknowledgments}

\bibliographystyle{apsrev}

\end{document}